\documentstyle[12pt]{article}
\begin{document}
\vspace{-2.6cm}

\title{\large\bfseries
PHYSICAL MODEL OF HADRONS\,:\,BARIONS and MESONS. \\
PHYSICAL ESSENCE of QUARKS and GLUONS AND \\
PHYSICAL INTERPRETATION OF THEIR PARAMETERS.}
\vspace{-2.4cm}

\author{\sffamily Josiph Mladenov Rangelov\,,\\
Institute of Solid State Physics\,-\, Bulgarian Academy of Sciences\,,\\
72\, blw\, Tzarigradsko chaussee\,,\,1784\,,\,Sofia\,,\,Bulgaria }

\date{}
\maketitle

\vspace{-1.2cm}

\begin{abstract}
 The physical model (PhsMdl) of the hadrons is offered by means of the obvious
analogy with the transparent surveyed PhsMdls of the vacuum and leptons in
our recent works. It is assumed that the vacuum is consistent by dynamides,
streamlined in junctions of some tight crystalline lattice. Every dynamide
contains a neutral pair of massless contrary point-like (PntLk) elementary
electric charges (ElmElc Chrgs): electrino $(-)$ and positrino $(+)$. By means
of the existent fundamental analogy between their properties and behaviour we
can understand the similarity and difference between them and assume that the
quark parameter aroma is determined by the value of its size of its circular
two-dimensional motion, while the quark parameter colour is determined by
orientation of the flat of the same circular two-dimensional motion in the
space. The colorless of the barions is explained by distribution of the same
circular two-dimensional motion of its elementary electric charge within
three mutually perpendicular flats. Then the exchange of the colors between
two quarks with different colors within some hadron can be interpretated as
some twisting of same hadron in the space. We give a new obvious physical
interpretation of the charge values of quarks, which gives some explanation
of angles of Cabibo and Weynberg. By some physical supposition about the
structure of charged intermediate vector bozon $W$ and uncharged intermediate
vector bozon $Z$ we have possibility to explain as the physical essence of
the strong, weak and electromagnetic interactions, so the outline of all
births, transformations and decays of the ElmPrts.
\end{abstract}

 Although up to the present nobody of scientists distinctly knows are there
some elementary micro particles (ElmMicrPrts) as a stone of the micro world
and what the elementary micro particle (ElmMicrPrt) means, there exists an
essential possibility for clear and obvious scientific consideration of the
unusual behaviour of the quantized micro particles (QntMicrPrts). It is well
known that the physical model (PhsMdl) presents at us as an actual ingradient
of every good physical theory. It may be used as for an obvious visual
teaching the occurred physical processes within the investigated phenomena,
so for doing with them a bigger capacity of its physical understanding and
mathematical description. All scientists think that the ElmMicrPrt display
itself as united and indivisible everywhere and always, but nobody demand for
 ElmMicrPrt to have no structure. About thirty years ago in the early theory
of the ElmMicrPrt some theoretical physicists assumed that all the ElmMicrPrts
have point-like (PntLk)-fashion and their behaviour are no dependent on the
presence of other ElmMicrPrt as neighbors. In further it was turned out that
the existence of the strong interaction between hadrons and of their own
magnetic dipole moment (MgnDplMmn) is caused by existence and exchange of the
virtual pions between them. Really, the strong interaction between nearest
hadrons is provided by the electromagnetic interaction, which is transferred
by frequent exchanging a virtual charged pion ($\pi^{\pm}$-meson). The virtual
charged pions ($\pi^{\pm}$-mesons), owing of its smallest mass between all
other mesons, are most probable for a frequent birth from all other virtual
ElmMicrPrts, which are caused by exchanged gluon within the quantized
electromagnetic field (QntElcMgnFld).

 We can assume without restriction that if the fine spread (FnSpr) elementary
electric charge (ElmElcChrg) of every charged lepton participates in
isotropic three-dimensional (IstThrDmn) relativistic quantized Schrodinger
(RltQntSchr) self-consistent strong-correlated (SlfCnsStrCrr) solitary
fermion vortical harmonic oscillations (SltFrmVrtHrmOscs) within its volume,
then the FnSpr ElmElcChrg of every charged hadron participates successive in
three anisotropic two-dimensional (AnIstTwoDmn) relativistic quantized
(RltQnt) self-consistent strong-correlated (SlfCnsStrCrr) solitary fermion
circular harmonic oscillations (SltFrmCrcHrmOscs) within its volume. If the
RltQntSchr SlfCnsStrCrr SltFrmVrtHrmOscs can be approximately described by the
orbital wave function (OrbWvFnc) of s-state ItsThrDmnNrlQnt boson harmonic
oscillator (BsnHrmOsc), the AnIstTwoDmn RltQnt SlfCnsStrCrr FrmCrcHrmOscs can
be approximately described by OrbWvFnc of p-state of AnIstThrDmn NrlQnt boson
harmonic oscillator (BsnHrmOsc). Therefore we can assume that if the averaged
FnSpr ElmElcChrg of charged lepton has one maximum, then the averaged FnSpr
ElmElcChrg of charged hadron has three maximums, which can be identified as
three FnSpr ElmElcChrgs of three quarks. But in a reality their is only one
FnSpr ElmElcChrg within every charged hadron, which is determined by electric
interaction only with the FnSpr ElmElcChrg of one maximum at a dispersion of
passing charged leptons. Indeed, at every dispersion of some FnSpr ElmElcChrg
of leptons from every hadron the FnSpr ElmElcChrg electromagnetically
interacts  (ElcMgnIntAct) only by one maximum of FnSpr ElmElcChrg of same
hadron.

 The existence of essential analogy between leptons and hadrons allows us to
assume that the quarks for the hadrons are analogous of the neutrinos for the
leptons. Therefore the physical means of the aroma of the quarks is analogous
of the physical means of the aroma of the neutrinos, which is determined by
the sizes of the vortical harmonic excitation of the neutral but fluctuating
vacuum (FlcVcm). Because of that we can suppose that if the neutrino is an
isotropic three dimensional solitary vortical excitation of a spherical
symmetry of the neutral vacuum, then the quark is a two-dimensional flat
circular soliton excitation of cylindrical symmetry of the same neutral
vacuum. But if all the neutrino are very stable excitation without elementary
electric charge (ElmElcChrg) and have unlimited time of live, all the quarks
are unstable without ElmElcChgr and have limited time of live. Moreover we
may assume also that if within some massive lepton (electron, muon
($\mu$-meson), tauon ($\tau$-meson)) its point-like (PntLk) ElmElcChrg takes
part simultaneously in three one-dimensional very strongly correlated
self-consistent fermion vortical harmonic oscillations along three axes,
orientired mutually perpendicular one to another in the space, then within
some massive charged hadron its PntLk ElmElcChrg or within other massive
,,uncharged'' hadron its two opportunity PntLk ElmElcChrgs, take part
successive in three two-dimensional very strongly correlated self-consistent
fermion circular harmonic oscillations along three planes, orientired
mutually perpendicular one to another in the space.  Therefore we may
naturally assume that three colours of the quark state correspond to three
planes of motion of the PntLk ElmElcChrg, orientired mutually perpendicular
one to another in the space.

 If the lepton state of its PntLk ElmElcChrg is determined by one spherical
state, describing the self- consist motion (Zitterbewegung) of its PntLk
ElmElcChrg by one s-OrbWvFnc, then the hadron state of its PntLk ElmElcChrg
is determined by three flat states, describing its three dimension motion,
composed by three flat cylindrical harmonic motions by three p-OrbWvFnc.
Moreover, each p-state correspond to one quark state with one of three
colours. Therefore the physical incomprehencible demand about colourless of
the hadrons obtains quite obvious and clear physical interpretation about the
demand of the spherical symmetry of three mutual perpendicular orientated
flats of two-dimensional fermion circular harmonic oscillations in our PhsMdl.
The change of the quark colour of the PntLk ElmElcChrg flat circular harmonic
motion occurs through emission or absorption of some gluon ($\delta$). When
the PntLk ElmElcChrg of some hadron, moving within quark of some colour,
absorbs some gluon (a quantized quasi-plane magnetic field), then this quarks
changes his colour and the hadron as whole twists the flat of the
two-dimensional fermion circular harmonic oscillations of his PntLk ElmElcChrg.

 In the beginning for an explanation of the existent experimental date some
theoretical physicists suppose that there are only three quarks and the same
number antiquarks. But for physical explanation of new experimental date which
are obtained during the carrying the following experiments, the scientists
had need from new three quarks and the same number antiquarks. There are a
common agreement that the existence of six quark (d,u,s,c,b,t) and the same
number antiquark ($\tilde d,\tilde u,\tilde s,\tilde c,\tilde b,\tilde t $)
is firmly determined by experimental research. Coming out from the supposed
symmetry between leptons and quarks some theoretical physicists very frequently
suppose a following list of their distribution on aroma:
\begin{eqnarray}\label{aa}
\left|\matrix {\nu_e & \nu_{\mu} & \nu_{\tau} \cr
e^{-} & \mu^{-} & \tau^{-} \cr u & c & t \cr
d & s & b } \right| \qquad \qquad
\left|\matrix {\tilde {\nu}_e & \tilde {\nu}_{\mu} & \tilde {\nu}_{\tau} \cr
 \tilde {e}^{-} & \tilde {\mu}^{-} & \tilde {\tau}^{-} \cr
 \tilde u & \tilde c & \tilde t \cr  \tilde d & \tilde s & \tilde b } \right|
 \end{eqnarray}

 Indeed, we can suppose that owing of the strong localization of the PntLk
ElmElcChrg of any massive hadron allow one to participate in very high
frequency oscillations within very small area of the space, what secures the
obtaining its QntElcMgnFld with very high electric and magnetic intensities
and very dense electromagnetic energy, which same FnSpr ElmElcChrg borrows
from the FlcVcm at its electromagnetic interaction with it by stochastic
exchanging the virtual photons (VrtPhtns). Then it is obviously that the
existence of the QntElcMgnFld with very high electric and magnetic intensities
and very dense electromagnetic energy can secures the need probability for
frequently births of virtual charged pions pairs within the immediate
neighborhood of the PntLk ElmElcChrg's position. Therefore we mast suppose
that the interacting hadrons are very quite near one to others. At these
conditions we may suppose also that when some hadron (for example some proton)
takes place in some nucleus and has as neighbors other hadrons with other
isotopic spin (neurtons) then during the time of the motion of its PntLk
ElmElcChrg within some $u$ quark without changing its color, the hadron is
visited by negative charged virtual pion ($\pi^{-}$-meson), emitted by
neighbor neutrons. Then both the free u-quark from the proton and the
u-antiquark ($\bar u$- quark) annihilated and negative PntLk ElmElcChrg
shifts from its d-quarks to the free d-quarks of the visited proton. After
some time some gluon from FlcVcm generates another pair of u-quark and
u-antiquark with another colour from the fluctuating vacuum (FlcVcm). Then
the negative PntLk ElmElcChrg shifts itself from d-quark of the visited
proton on the new born u-antiquark, which together with the d-quark of the
old transferring negative charged pion ($\pi^{-}$-meson) build a new negative
charged pion. In the time of such a visit the positive PntLk ElmElcChrg
uninterruptedly participated within the flat two-dimensional fermion circular
harmonic oscillations within the u-quark of the visited proton. After the
departure of the negative PntLk ElmElcChrg from the new neutron the positive
PntLk ElmElcChrg shifts into the new born u-quark, which have the colour,
different from the colour of old quark, from which it left. After such an
exchange twisted proton is ready for a new visit from another negative
virtual pion ($\pi^{-}$- meson). In such a natural way the exchange of the
negative virtual pion between some pair of proton and neutron causes the
existence of the strong interaction.

 If the three different aroma of leptons correspond to three different sizes
of the isotropic three dimensional relativistic quantized (IstThrDmnRltQnt)
Schrodinger self-consistent strongly correlated (SchrSlfCnsStrCrr) solitary
fermion vortical harmonic oscillations (SltFrmVrtHrmOscs) of its FnSpr
ElmElcChrg, composed by three SlfCnsStrCrr one-dimensional relativistic
quantized harmonic oscillations, then the three different aroma of hadron
quark state correspond to three sizes of three two-dimensional relativistic
quantized fermion circular harmonic oscillations of its FnSpr ElmElcChrg.
Therefore we may suppose that if within the massive lepton its averaged (over
spread (OvrSpr)) ElmElcChrg has one maximum only, then within the massive
hadron its averaged OvrSpr ElmElcChrg has three maximums only, each of which
may be identified by us as the ElcChrg of the quark or parton.  The exchange
of the FnSpr ElmElcChrg of its quark colour state occurs through absorption
or emission of some gluon ($\delta$).

 The gluons are hypothetical particle, which exchange between two quarks with
different colors (space orientations) achieve the strong interaction between
them. It is turn out that they have not a ElmElcChrg and rest mass, but have
a proper mechanical moment (spin) and quantized magnetic field with a proper
magnetic dipole moment (MgnDplMmn). Therefore in magnetic interaction
(MgnIntAct) between the QntMgnFlds of the gluon and quarks ensures the turn
of space orientation (change of color) of gluon. It is turn that the impulse
(mechanical momentum) sum of all gluons within some hadron is equal of the
impulse sum of all quarks within same hadron. As the gluon spin is 1$\hbar$,
therefore at its habitual virtual decay within hadron one can turn into pair
virtual quark and antiquark with paralel spins. But such a virtual pair is no
virtual meson, as more mesons with small mass have a spin with a zero value.
Therefore such a decay of some gluon into a pair of some virtual quark and
antiquark could effectively turn upsite-down the spin of one of meson quark
and in such a way enssure the decay of this meson in first into charged
intermediate vector bozon $W$ and in second into massive and massless leptons
of equal aromas.  The behavior of eight gluons can be described by eight
Gell-Man's mathrices of SU(3) space.

 We must note that the PntLk ElmElcChrg of the charged admeson is moving
successively in two opposite orientated flats.  As in positive $\pi$-meson
its positive PntLk ElmElcChrg (positrino) is successively moving in an u-quark
state and in a $\tilde{d}$-antiquark state, which have an opposite parallel
orientations, then in negative $\pi$-meson its negative PntLk ElmElcChrg
(electrino) is successively moving in a d-quark state and in a $\tilde{u}$
-antiquark state, which have opposite parallel orientations. In same way in
the positive K-meson its positive massless PntLk ElmElcChrg (positrino) is
successively moving in an u quark state and in a $\tilde{s}$-antiquark state,
which have opposite parallel orientations, and in the negative K-meson its
negative PntLk ElmElcChrg (electrino) is successively moving in a s-quark
state and in a $\tilde{u}$-antiquark state, which have opposite parallel
orientations.  However in the neutral admesons two opposite charged PntLk
ElmElcChrgs (electrino and positrino) are moving in four flats, which have in
pair opposite parallel orientations, where are moving separately two opposite
charged PntLk ElmElcChrgs.  For instance in $\pi^o$-meson its massless
electrino is moving in d-quark state and in $\tilde{u}$-antiquark state,
which both have opposite parallel orientations, and its positrino is moving
in u quark state and in $\tilde{d}$ antiquark state, which both have opposite
parallel orientations, mutually perpendicular orientated to the first pair
flats. In such a fashion in $K^o$ -meson its electrino is moving in a d quark
state and in a $\tilde{u}$-antiquark state, which both have opposite parallel
orientations, and its positrino is moving in an u quark state and in
$\tilde{s}$ antiquark state, which both have opposite parallel orientations,
but the second are mutual perpendicular oriented to the first pair flats of
moving electrino. In such a fashion in $\tilde{K}^o$- meson its electrino is
moving in a s-quark state and in a $\tilde{u}$ antiquark state, which both
have opposite parallel orientations, and its positrino is moving in an u quark
state and in a $\tilde{d} $-antiquark state, which both have opposite parallel
orientations, but the second are mutual perpendicular oriented to first pair
flats of moving electrino. Owing of a small difference between the
compositions of quarks, from which are built the $K^o$-meson and the
$\tilde{K}^o$-meson the some fluctuations in the FlcVcm have energy, enough
for their conversion one meson in its antimeson.
\begin{eqnarray}\label{au1}
K^o (\,d\,\uparrow\,,\,\tilde{s}\,\downarrow\,) =
(\,u\,\downarrow\,,\,W^-\,\uparrow\,,\,\tilde{s}\,\downarrow\,) =
(\,u\,\downarrow\,,\,\tilde{c}\,\uparrow\,) =
(\,s\,\uparrow\,,\,W^+\,\downarrow\,,\,\tilde{c}\,\uparrow\,) =
\tilde{K}^o (\,s\,\uparrow\,,\,\tilde{d}\,\downarrow\,)
\end{eqnarray}

 In a result of such mutual conversions are obtained short-lived and
long-lived $K^o$-mesons with different parameters and scheme of decay.
\begin{eqnarray}\label{av}
K^o_l = \frac{( K^o + \tilde{K}^o )}{\sqrt{2}} \quad {\bf and} \quad
K^o_s = \frac{( K^o - \tilde{K}^o )}{\sqrt{2}}
\end{eqnarray}

The similar fashion of the meson construction we may find in the $D^o$-mesons,
where its positive PntLk ElmElcChrg is successively moving in an u-quark
state and in a $\tilde{d}$-antiquark state, which both have opposite parallel
orientations, and its negative PntLk ElmElcChrg is successively moving in a
d-quark state and in a $\tilde{c}$-antiquark state, which both have opposite
parallel orientations, but the second are mutually perpendicular to first pair
flats. In such a fashion in $\tilde{D}^o$-meson its electrino is moving in a
d-quark state and in a $\tilde{u}$-antiquark state, which both have opposite
parallel orientations, and its positrino is moving in a c quark state and in
a $\tilde{d}$-antiquark state, which both have opposite parallel orientations,
but the second are mutually perpendicular to first pair flats.
\begin{eqnarray}\label{au2}
D^o (\,u\,\uparrow\,,\,\tilde{c}\,\downarrow\,) =
(\,d\,\downarrow\,,\,W^+\,\uparrow\,,\,\tilde{c}\,\downarrow\,) =
(\,d\,\downarrow\,,\,\tilde{s}\,\uparrow\,) =
(\,c\,\uparrow\,,\,W^-\,\downarrow\,,\tilde{s}\,\uparrow\,) =
=\tilde{D}^o (\,c\,\uparrow\,,\,\tilde{u}\,\downarrow\,)
\end{eqnarray}

  As for the radiation of the spontaneous real photon (RlPhtn) from the
excitative atom it is necessary the presence of the virtual photon (VrtPhtn)
for the creation of the electric dipole moment (ElcDplMm), so for the decay
of a charged $\pi$-meson it is necessary the presence of a virtual gluon for
an overturning of the spin of one of its quarks, by which charged $\pi$-meson
turns into charged virtual $\rho$-meson, which can immediately decay into
charged intermediate vector boson $W$. At the subsequent transfer of the
in a pair of massive and massless leptons of equal aroma the participating in
the decay gluon go back in the FlcVcm. Indeed, I think that instead of the
equations of the incomprehencible decay :
\begin{eqnarray}\label{a}
\pi^+ \quad\Longrightarrow\quad W^+ \quad \Longrightarrow\quad \mu^+ \,+ \,
\nu_\mu\quad,
\end{eqnarray}

\vspace{-0.7cm}
\begin{eqnarray}\label{b}
\pi^- \quad\Longrightarrow\quad W^- \quad
\Longrightarrow\quad \mu^- \,+\,\tilde{\nu}_\mu \quad,
\end{eqnarray}

\vspace{-0.7cm}
\begin{eqnarray}\label{c}
\pi^+ \quad\Longrightarrow\quad W^+ \quad
\Longrightarrow\quad e^+ \,+\,\nu_e \quad,
\end{eqnarray}

\vspace{-0.7cm}
\begin{eqnarray}\label{d}
\pi^- \quad\Longrightarrow\quad W^- \quad
\Longrightarrow\quad e^- \,+\,\tilde{\nu}_e \quad,
\end{eqnarray}

\vspace{-0.7cm}
\begin{eqnarray}\label{e}
\pi^o \quad\Longrightarrow\quad W^+ \,+\, W^- \quad
\Longrightarrow\quad\gamma\,+\,\gamma\quad,
\end{eqnarray}

\vspace{-0.7cm}
\begin{eqnarray}\label{f}
\pi^o \quad\Longrightarrow\quad W^+ \,+\, W^- \quad
\Longrightarrow\quad\gamma \,+\, e^+ \,+\, e^- \quad,
\end{eqnarray}

we must used the following equations :
\begin{eqnarray}\label{g}
\pi^+ \,+\,\delta\quad\Longrightarrow\quad W^+ \quad
\Longrightarrow\quad \mu^+ \,+\,\nu_\mu \quad,
\end{eqnarray}

\vspace{-0.7cm}
\begin{eqnarray}\label{h}
\pi^- \,+\,\delta\quad\Longrightarrow\quad W^- \quad
\Longrightarrow\quad \mu^- \,+\,\tilde{\nu}_\mu \quad,
\end{eqnarray}

\vspace{-0.7cm}
\begin{eqnarray}\label{i}
\pi^+ \,+\,\delta\quad\Longrightarrow\quad W^+ \quad
\Longrightarrow\quad e^+ \,+\,\nu_e \quad,
\end{eqnarray}

\vspace{-0.7cm}
\begin{eqnarray}\label{j}
\pi^- \,+\,\delta\quad\Longrightarrow\quad W^- \quad
\Longrightarrow\quad e^- \,+\,\tilde{\nu}_e \quad,
\end{eqnarray}

\vspace{-0.7cm}
\begin{eqnarray}\label{k}
\pi^o \quad\Longrightarrow\quad W^+ \,+\, W^- \quad
\Longrightarrow\quad\gamma \,+\,\gamma \quad,
\end{eqnarray}

\vspace{-0.7cm}
\begin{eqnarray}\label{l}
\pi^o \quad\Longrightarrow\quad W^+ \,+\, W^- \quad
\Longrightarrow\quad\gamma\,+\, e^+ \,+\, e^- \quad,
\end{eqnarray}

 As it is easy to see from eqns.$(\ref{k})$ and $(\ref{l})$ that there is no
necessity for participating some gluon $\delta$ in decay of the $\pi^o$-meson.
For certain of that the half-life time ($\tau_\pi^+\,=\,2.6\,\times\,10^-8
s$) of the charged $\pi$-mesons is very different from the half-life time
($\tau_\pi^o\,=\,8.3\,\times\,10^-17 s$) of the neutral $\pi$-mesons.

 It seems to me the existance of two very interesting facts, having common
physical cause.The first is concurrence of the energy of one degree of
freedom in charged lepton $\mu$-meson and in charged admeson $\pi$-meson.
Indeed, if in isotropic three dimensional solitary vortical harmonic
ocsillations of FnSpr ElmElcChrg of $\mu$-meson have three degrees of freedom
and therefore $3\,\hbar\,\omega = 2\,m\,C^2 = 213.2$Mev. Hence the energy of
one degree of freedom can be determined $\frac{\hbar\,\omega}{2} = 35.5 $Mev.
If we take into consideration that the FnSpr ElmElcChrg of $\pi$-meson takes
participation in two quasi-plane circular harmonic oscillations with
opportunity orientations and therefore has energy 2$\hbar\omega = 139.6$Mev.
Hence the energy of one degree of freedom can be determined $\frac{\hbar\,
\omega}{2} = 34.9 $Mev. As we can see by comparision of two results this
coincidence is very accurate. On this reason we can assume that the areas of
their oscillations must also coincidence and therefore the OrbWvFnc of both
FnSpr ElmElcChrg. May be therefore the decay of the positive (negative)
charged $\pi$-meson in $100\%$ occurs through the positive (negative)
$\mu$-meson and $\mu$-neutrino (antineutrino) as we can see from (\ref{a},
\ref{b}, \ref{g}, \ref{h}). This second coincidence gives us many correct
answer of the question for inner structure of the elementary micro particles
(ElmMicrPrts).

 It is useful to remember that the existence of some analogy between leptons
and hadrons allows us to assume that the quarks for the hadrons are analogous
of the neutrinos for the leptons. Therefore the physical means of the aroma
of the quarks is analogous of the physical means of the aroma of the
neutrinos, which determines the sizes of the type excitation of the neutral
but fluctuating vacuum (FlcVcm).  Because of that we can suppose that if the
neutrino is three dimensional spherical soliton seclusion excitation of the
neutral vacuum, then the quark is the sum of two two-dimensional flat soliton
excitation of same neutral vacuum. As the vacuum is formed by neutral dynamides,
which are made from two contrary massless electric charges ((-) electrino and
(+) positrino) through their dense streamlining, then its solution seclusion
oscillation secures the joint motion of the oscillating FlcVcm and the PntLk
ElmElcChrg. However if the neutrino can accept the PntLk ElmElcChrg for an
unlimited time at a creation of the massive lepton, then the quark can accept
the PntLr ElmElcChrg only for a limited time t or 2t at every its visit, after
which the PntLk ElmElcChrg must leave it and go on to another quark. In this
a way every PntLk ElmElcChrg in every hadron continuously and alternately
changes the quarks of the hadron, in which it temporary pass into.  Therefore
as we could see the average distribution of the spread electric charge, we
think that the electric charge of one type quark is $\frac{1e}{3}$ and the
electric charge of an other type quark is $\frac{2e}{3}$. The time of pass
$\tau$ or $2\tau$ of the PntLk ElmElcChrg within some quark area is determined
in depending on the kind (symmetric or antisymmetric) of the neutral flat
vortical oscillation of the FlcVcm. So any flat oscillation can be presented
as a sum of two different flat oscillations, orientated in opposite directions.
Therefore two kind of the quarks could be presented as symmetrical and
antisymmetrical sums of two one-sided orientated protoquarks $\left|PK\right|$
and other-sided orientated antiprotoquark $\left|APK\right|$ :
\begin{eqnarray}\label{m}
\left|\matrix{SC \cr ASC}\right| \,=
\, \left|\matrix{\cos(\theta_o) & \sin(\theta_o) \cr
 -\sin(\theta_o) & \cos(\theta_o)}\right|\,\times
\,\left|\matrix{PQ \cr APQ}\right| \,
\end{eqnarray}

where the $\left|SC\right|$ means a symmetric combination of one-side flat
oscillation of protoquark $\left|PQ\right|$ and of one-side flat oscillation
of antiprotoquark $\left|APQ\right|$ , while the $\left|ASC\right|$ means an
antisymmertic combination of one-side flat oscillation of protoquark
$\left|PQ\right|$ and of one-side flat oscillation of antiprotoquark
$\left|APQ\right|$. If we assume that $ \theta_o\,=\,30^o $ then
$\sin(\theta_o)\,=\,\frac{1}{2}$ and $\cos(\theta_o)\,=\,\frac{\sqrt{3}}{2}$.
Therefore in this case we can obtain from $(\ref{m})$ the following equation :
\begin{eqnarray}\label{n}
\left|\matrix{SC \cr ASC}\right| \,=
\,\left|\matrix{\frac{\sqrt{3}}{2} & \frac{1}{2} \cr
 -\frac{1}{2} & \frac{\sqrt{3}}{2}}\right|\,\times
\,\left|\matrix{PQ \cr APQ}\right|\,
\end{eqnarray}

Hence the angle $\theta_o$ determine the ratio of the participation of two
oscillations. This interpretations help us to understand the physical mean of
number $\frac{1}{3}$ and $\frac{2}{3}$. Indeed, if the $\left|SC\right|$
means the symmetric combination of two opposite orientated oscillations then
the probability to find the PntLk ElcChrg in this state is determined by the
difference of the quadrats of its amplitudes $\frac{3}{4}$ and $\frac{1}{4}$.
By this natural way we have obtain the number $\frac{2}{4}\,=\,\frac{1}{2}$.
Then, if the $ \left|ASC\right|$ means the antisymmetric combination of two
opposite orientated oscillations then the probability to find the PntLk
ElcChrg in this state is determined by sum of the quadrats of its amplitudes
$\frac{3}{4}$ and $\frac{1}{4}$. By this natural way we have obtain the number
$\frac{4}{4}\,=\,1$. Consequently the number $\frac{1}{3}$ and the number
$\frac{2}{3}$ mean that the probability of finding the PntLk ElmElcChrg in
the state of the SC is $\frac{1}{3}$ and the probability of finding the PntLk
ElmElcChrg in the state of the ASC is $\frac{2}{3}$ . Therefore we think that
the values of the electric charges are $\frac{1e}{3}$ and $\frac{2e}{3}$.

 By means of this our interpretation of our PhsMdl we can obtain the real
values of the Weinberg angle and of the Cabibo angle. Indeed, in the
probability of the finding the PntLk ElmElcChrg in the $d$ quark state is
equal of $\frac{1}{3}$ and the probability of the finding same PntLk
ElmElcChrg in the $u$ quark state is equal of $\frac{2}{3}$, then it is
obviously clearly that the transition probability from one kind of the quark
to another kind of quark is equal of their product. Therefore we have obtained
that ${\sin(\theta_w)}^2 = \frac{2}{9}$ and $\sin(\theta_w) = \frac{\sqrt{2}}
{3}$. Consequently, $\theta_w = \arcsin{\frac{\sqrt{2}}{3}}\,$ or
$\theta_w = 28^o\,8^{,}\,$.

 In such a naturally easy, physically substantiated and mathematically correct
way we have achieved Weinberg's angle values, which very well coincides with
its value, determined by the experiment. It is very important that the value,
$\tilde{\theta}_w\,=\,38^o\,$, which is determined by means of the $SU(2)\,
\times\,SU(1)$ group method, is very different from the experimental value.

 We have explained that besides the colour the quark state has also the aroma,
which difference ensures perfectly different properties of the quark state.
We have supposed, that three different aroma of the quark correspond to three
sizes of the two-dimensional flat fermion harmonic oscillations of its PntLk
ElmElcChrg. It turns out that there are possibility enough to give us clear
physical interpretation and mathematical substantiation for obtaining the
Cabibo angle value, determined by experiment. Indeed, we may assume, that in
a result of the different cover of the OrbWvFncs of different aroma of quarks,
there will be different possibility for a transition of one aroma of quark in
another aroma of quark with same ClcChrg. Some times this possibility is
interpreted as a reflection of some mutual influence between them. Therefore
we may assume that each quark is consists from quarks of different aroma but
of same ElcChrg. In this supposing we may roughly write the following
equations :
\begin{equation}\label{o1}
\,d_1\,=\,d_o\,\cos{\alpha}\,+\,s_o\,\sin{\alpha}\,;
\,s_1\,=\,s_o\,\cos{\alpha}\,-\,d_o\,\sin{\alpha}\,;
\end{equation}

\vspace{-0.6cm}

\begin{equation}\label{o2}
\,u_1\,=\,u_o\,\cos{\beta}\,+\,c_o\,\sin{\beta}\,;
\,c_1\,=\,c_o\,\cos{\beta}\,-\,u_o\,\sin{\beta}\,;
\end{equation}

 It is easy to verify that :
\begin{equation}\label{o3}
\,\tilde{u}_1\,d_1\,+\,\tilde{c}_1\,s_1\,=
\,(\tilde{u}_o\,d_o\,+\,\tilde{c}_o\,s_o\,)\,\cos(\alpha - \beta)\,+
\,(\tilde{u}_o\,s_o\,-\,\tilde{c}_o\,d_o\,)\,\sin(\alpha - \beta)\,;
\end{equation}

\vspace{-0.6cm}

\begin{equation}\label{o4}
\,\tilde{u}_1\,s_1\,-\,\tilde{c}_1\,d_1\,=
\,(\tilde{u}_o\,s_o\,-\,\tilde{c}_o\,d_o\,)\,\cos(\alpha - \beta)\,-
\,(\tilde{u}_o\,d_o\,+\,\tilde{c}_o\,s_o\,)\,\sin(\alpha - \beta)\,;
\end{equation}

 If we present the difference $(\alpha\,-\,\beta)\,$ as Cabibo angle $\theta_c$
then from eqns.$~(\ref{o3})$ and $(\ref{o4})$ one becomes clear that at the
 participation of the quarks of equal aroma $(\tilde{u}_1\,d_1\,+
\,\tilde{c}_1\,s_1)$ and different aroma
$(\tilde{u}_1\,s_1\,-\,\tilde{c}_1\,d_1)$ in the weak interaction reaction,
then Cabibo angle determines the decay probabilities with the hypercharge
change and hypercharge conservation.  Therefore Cabibo angle value is
determined by equation $\sin{\theta_c}\,=\,\frac{2}{9}$ and
$\theta_c\,=\,\arcsin{\frac{2}{9}}\,=\,12^o\,50^{,}$. In such a naturally,
 physically clear and mathematically substantiated way we have easy achieved
Cabibo angle value $12.84^o\,$ which very well coincides with the value
$12.7^o\,$, determined by the experiment.

 We assume that the proton is composed from one free positive PntLk ElmElcChrg,
which visits consecutively two states of $u$-quark.The free $u$-quark and the
$\tilde d$-antiquark are frequently visited by negative PntLk ElmElcChrg of
virtual negative charged $\pi^{-}$-mesons and the neutron is composed from
two free opposite charged PntLk ElmElcChrgs, the positive one is oscillating
in a state of $u$-quark and the negative one is oscillating in a state of two
$ d $-quarks. Very frequently the negative PntLk ElmElcChrg after forming the
virtual negative charged point ($\pi^{-}$-meson) by means of birthed pair of
$u$-quark and $\tilde u$-antiquark owing of an absorption of some virtual gluon.
The birthed by this way virtual negative pion can participate in strong
interaction between very close neighbor proton and neutron by its exchanging.

 However I have possibility to show and you to understand by means of some
decay relation that negative charged intermediate vectorial boson  $W^{-}$
has spin minus $\hbar $,and the positive charged intermediate vectorial boson
$W^{+}$ have a spin $\hbar$, while the neutral intermediate vectorial boson
$Z^{o}$ have a spin zero $\hbar$. Therefore the PntLk ElmElcChrg during the
exchange of its self-consistent motions within one ElmMicrPrt and within other
ElmMcrPrt transfers in a form of charged intermediate vectorial boson $W$.
Indeed, the observing the law of total spin conservation at all decay with
participating the weak interaction, when the PntLk ElmElcChrg takes the form
of the charged intermediate vector boson $W^{\pm}$, causes fulfillment of some
strong selection rules from all ElmMicrPrt. The consideration of spins of the
quarks and charged intermediate vector bozon $ W $ allow us to understand why
the positive charged intermediate vectorial boson $ W^{+}$ emits only right
quarks and selects only left quarks for participating with them in a weak
interacting, while the negative charged intermediate vectorial boson $ W^{-}$
emits only left quarks and selects only right quarks for participating with
them in a weak interaction. The neutral intermediate vector boson $ Z^o $ don't
select the polarization of the quarks with which it participate in the weak
interaction.

 The weak interaction performs an important role in the relation between
hadrons and leptons as it is display in nuclear $\beta$ decay. Indeed, it is
possible some a negative PntLk ElmElcChrg (electrino) of any neutron, which
takes part within the oscillations of $d$-quark, to change its self-consistent
motion and moves on the oscillations of $\tilde u$-quark, which are unstable
and therefore it decay in a negative charged intermediate vector boson $W^{-}$
and $u$-quark. In such a naturally way the neutron transforms itself in proton
and the residual the negative charged intermediate vector boson $W^{-}$ after
its incorporation with some new created electron neutrino $ \nu_e $ form the
electron, while the other new created antineutrino $\tilde \nu_e $ go a free
away. When some positive PntLk ElmElcChrg (positrino) of any proton, which
takes part in the oscillations of the $u$-quark, changes his self-consistent
motion and moves on the oscillations of the $\tilde{d}$-quark, which are
unstable and therefore it decay in positive charged intermediate vector boson
$ W^{+}$ and $d$-quark. In such a naturally way the proton transforms itself
in a neutron and the residual the positive charged intermediate vector boson
$ W^{+}$ after its incorporation with some new created electron antineutrino
$ \tilde \nu_e $ form the positron, while the other new created neutrino
$\nu_e$ go a free away. Therefore the described $\beta$-transitions can be
written in the following form :
\begin{equation}\label{p}
n(u,d,d)\,=\,p(u,u,d)\,+\,W^-\,=\,p(u,u,d)\,+\,e^-\,+\,\tilde{\nu}_e\,
\end{equation}

\vspace{-0.6cm}

\begin{equation}\label{q}
p(u,u,d)\,=\,n(u,d,d)\,+\,W^+\,=\,n(u,d,d)\,+\,e^+\,+\,\nu_e\,
\end{equation}

As within the areas of every nucleon there are two virtual opposite charged
$\pi$-mesons, at decreasing the distance between nucleons begins the creation
of some correlation between their motions, which creates some decreasing of
their total energy. At a rather decreasing of the distance between the nucleons
begins some collectivization of all virtual charged $\pi$-meson. In this way
we understand that internucleonic forces are analogous of interatomic Van-der-
Waals force between neutral atoms.

 Our PhsMdls explain as the structure of hadrons and the nature of their
interaction so the existence of a possibility for joint description of a
field and substantial form of the matter as unity whole in the physical
science, which are submitted to an united, fundamental and invariable laws of
nature.

 As all existent ElmMicrPrts are excitement of the vacuum then all of them
will move freely through one without any damping, that is to say without to
feel the existence of the vacuum. Moreover, the existence of some McrPrt in
the vacuum twists its cristalline lattice.This twist of the neutral vacuum
excites the gravitation field of the ElmMicrPrt's mass, which will influence
by using of some force upon mass of another ElmMicrPrt and upon its behaviour.

  We can understand the physical essence of the hadrons and their structure
and characteristics on these obvious representations. Our physical model
explains as the structure of hadrons and the nature of their three type of
interactions, so the existence of the possibility for a joint description of
a field and  substantial form of the matter as unity whole in the physical
science, which  are submitted to an united, fundamental and invariable laws
of nature. By this natural way we can see the unity of the field charged and
neutral excitations within the vacuum and its substantial charged excitations,
offered by modernity relativistic quantum mechanics (RltQntMch), quantum
electrodynamics (QntElcDnm) and quantum theory of field (QntThrFld).

 I cherish the hope that this consideration of my physical model of the
hadrons from my new point of view will be of great interest for all scientists.
My very quality and good interpretation of behaviour and structure of hadrons
and their interactions with its corresponding experimentally determined values
gives us the hope for correctness of our beautiful, simple and preposterous
physical model of hadrons and fine,extraordinary ideas,which have been
inserted  at its construction.

\end{document}